\documentclass[letter]{aa}  

\usepackage{graphicx}
\usepackage{txfonts}
\usepackage{xspace}
\usepackage{color}
\usepackage{booktabs}

\usepackage{environ}
\NewEnviron{splitequation}{%
    \begin{equation}\begin{split}
        \BODY
    \end{split}\end{equation}}

\newcommand{\software}[1]{\texttt{#1}}

\begin{document}

   \title{Cosmology from large-scale structure}

   \subtitle{Constraining $\Lambda$CDM with BOSS}

   \author{Tilman~Tröster,\inst{1}\thanks{\email{ttr@roe.ac.uk}}
          Ariel.~G.~Sánchez,\inst{2}
          Marika~Asgari,\inst{1}
          Chris~Blake,\inst{3}
          Mart\'in Crocce,\inst{4}
          Catherine~Heymans,\inst{1,5}
          Hendrik~Hildebrandt,\inst{5}
          Benjamin~Joachimi,\inst{6}
          Shahab~Joudaki,\inst{7}
          Arun~Kannawadi,\inst{8}
          Chieh-An~Lin\inst{1}
          \and
          Angus~Wright\inst{5}
                 }
    \authorrunning{T.~Tröster et al.}

   \institute{Institute for Astronomy, University of Edinburgh, Royal Observatory, Blackford Hill, Edinburgh, EH9 3HJ, UK
         \and
          Max-Planck-Institut f\"ur extraterrestrische Physik, Postfach 1312, Giessenbachstr., 85741 Garching, Germany
         \and
         Centre for Astrophysics \& Supercomputing, Swinburne University of Technology, P.O. Box 218, Hawthorn, VIC 3122, Australia
         \and
         Institut de Ci\'ences de l’Espai, IEEC-CSIC, Campus UAB, Carrer de Can Magrans, s/n, 08193 Bellaterra, Barcelona, Spain
         \and
            German Centre for Cosmological Lensing, Astronomisches Institut, Ruhr-Universität Bochum, Universitätsstr. 150, 44801, Bochum, Germany
          \and
          Department of Physics and Astronomy, University College London, Gower Street, London WC1E 6BT, UK
          \and
          Department of Physics, University of Oxford, Denys Wilkinson Building, Keble Road, Oxford OX1 3RH, UK
          \and
          Leiden Observatory, Leiden University, P.O.Box 9513, 2300RA Leiden, The Netherlands
}

   \date{Received September 19, 2019; accepted December 18, 2019}

\abstract{We reanalyse the anisotropic galaxy clustering measurement from the Baryon Oscillation Spectroscopic Survey (BOSS), demonstrating that using the full shape information provides cosmological constraints that are comparable to other low-redshift probes. 
We find $\Omega_\mathrm{m} = 0.317^{+0.015}_{-0.019}$, $\sigma_8 = 0.710\pm 0.049$, and $h = 0.704\pm 0.024$ for flat $\Lambda$CDM cosmologies using uninformative priors on $\Omega_\mathrm{c}h^2$, $100\theta_\mathrm{MC}$, $\ln 10^{10} A_{s}$, and $n_{s}$, and a prior on $\Omega_\mathrm{b}h^2$ that is much wider than current constraints.
We quantify the agreement between the Planck 2018 constraints from the cosmic microwave background and BOSS, finding the two data sets to be consistent within a flat $\Lambda$CDM cosmology using the Bayes factor as well as the prior-insensitive suspiciousness statistic.
Combining two low-redshift probes, we jointly analyse the clustering of BOSS galaxies with weak lensing measurements from the Kilo-Degree Survey (KV450). 
The combination of BOSS and KV450 improves the measurement by up to 45\%, constraining $\sigma_8 = 0.702\pm 0.029$ and $S_8 = \sigma_8\sqrt{\Omega_\mathrm{m}/0.3} = 0.728\pm 0.026$. 
Over the full 5D parameter space, the odds in favour of a single cosmology describing galaxy clustering, lensing, and the cosmic microwave background are {$7\pm2$}. 
The suspiciousness statistic signals a {$2.1\pm0.3\sigma$} tension between the combined low-redshift probes and measurements from the cosmic microwave background.

}

   \keywords{large-scale structure of the Universe -- cosmological parameters
               }

   \maketitle
%

\section{Introduction}
The last decade has seen the field of cosmology being transformed into a precision science, with many of the parameters that describe our Universe being constrained to sub per-cent precision.
This remarkable achievement has been largely driven by the observations of the cosmic microwave background (CMB) conducted by the WMAP \citep{WMAP9} and Planck \citep{Planck2018-cosmology} satellites.
While the constraining power of the CMB still reigns supreme, other, independent, observations of the more recent Universe have begun to be able to constrain certain parameters at a precision comparable to that achieved by Planck \citep[e.g.,][]{DES-3x2pt,Riess2019}.
This has led to the rise of a range of `tensions' between data sets: disagreements that do not reach the level of statistical significance to warrant a claim to the detection of deviation from $\Lambda$CDM but that are large enough to cause discomfort because their occurrences are deemed to be somewhat too unlikely to be a statistical fluke.

In this \textit{Letter}, we provide another datum in this evolving picture of cosmic concordance by providing new, independent constraints on $\Lambda$CDM from the clustering of galaxies. 
One of the most powerful probes of cosmology in the low-redshift Universe comes from observations of the large-scale structure (LSS) of the Universe.
Analyses of the clustering of galaxies, either through measurements of the baryon acoustic oscillations (BAO), redshift-space distortions (RSD), or the full shape of two-point statistics by the Baryon Oscillation Spectroscopic Survey (BOSS) collaboration \citep{Alam2017}, have been able to break degeneracies in the parameter space allowed by Planck, thus further increasing the precision of the parameters that underlie the $\Lambda$CDM concordance model of cosmology and ruling out deviations from it. 
These analyses only constrained $\Lambda$CDM, or extensions thereof, in conjunction with other data sets and did not attempt to constrain $\Lambda$CDM with BOSS data alone. 
Instead, the consensus analysis of the final BOSS Data Release 12 (DR12) data \citep{Alam2017} provides constraints in terms of geometric quantities describing the tangential and radial BAO scales, as well as the growth rate of structure and amplitude of matter fluctuations, $f\sigma_8$. 
In this parameterisation, a particular point in parameter space need not correspond to a valid $\Lambda$CDM cosmology, since the different distance measures and growth of structure are considered to be independent. 
Full-shape analyses of the anisotropic clustering signal of galaxies are able to break degeneracies \citep[][]{Loureiro2019, Kobayashi2019} between parameters and thus constrain cosmology without relying on external data sets.

In this \textit{Letter}, we revisit the full-shape analysis of correlation function wedges of \citet[][hereafter S17]{Sanchez2017} and derive constraints on the parameters of flat $\Lambda$CDM cosmologies.
In Sect.~\ref{sec:methods}, we review the methodology and data used in S17 and comment on the changes and additional model validation carried out for the present analysis. 
Section~\ref{sec:results} presents the constraints on $\Lambda$CDM that we can derive from the clustering of BOSS galaxies, while Sect.~\ref{sec:discussion} discusses these results, both by themselves, and in the context of other low-redshift cosmological probes. 
Specifically, we perform a joint analysis with cosmic shear measurements from the Kilo-Degree Survey  \citep[KV450,][]{Hildebrandt2018}  to showcase the power such combined probe studies will gain over the next decade.
Finally, we conclude in Sect.~\ref{sec:conclusions}.

\section{Methods}
\label{sec:methods}
This work closely follows the analysis of S17, only changing the sampling space and priors. 
In this section we briefly review the data and modelling and refer the interested reader to S17 for details.

\subsection{Data}
We consider the full BOSS DR12 data set, which is split into two redshift bins $0.2 \leq z <0.5$ and $0.5 \leq z < 0.75$ \citep[see][]{Alam2017}. 
The redshifts are converted into distances at a fiducial cosmology with $\Omega_\mathrm{m}=0.31$ and $h=0.7$. 
For both redshift bins we measure the anisotropic correlation function $\xi(\mu, s)$ using the \citet{Landy1993} estimator, where $\mu$ is the cosine of the angle between the line of sight and the separation vector  between the pair of galaxies, and $s$ denotes the comoving distance between the pair of galaxies.  
The correlation functions are then binned in $\mu$ into three equal-sized `wedges': $0 \leq \mu < 1/3$, $1/3 \leq \mu < 2/3$, $2/3 \leq \mu < 1$; and binned in $s$ into bins of width $\Delta s = 5\,h^{-1}\mathrm{Mpc}$ between $s_{\rm min}=20\,h^{-1}\mathrm{Mpc}$ and $s_{\rm max}=160\,h^{-1}\mathrm{Mpc}$. 
The data covariance matrix is estimated from 2045 \software{MD-Patchy} mock catalogues \citep{Kitaura2016}.

\subsection{Model}
\label{sec:model}
The non-linear evolution of the matter density is described by a formulation of renormalised perturbation theory \citep{Crocce2006} that restores Galilean invariance, referred to as gRPT (Crocce et al. in prep.). 
The galaxy density $\delta_{g}$ is related to the matter density $\delta$ by \citep{Chan2012}
\begin{equation}
\label{equ:bias-model}
	\delta_{g} = b_{1}\delta + \frac{b_{2}}{2}\delta^{2} + \gamma_{2}\mathcal{G}_{2} + \gamma_{3}^{-}\Delta_{3}\mathcal{G} + \dots \ .
\end{equation}
The operators $\mathcal{G}_{2}$ and $\Delta_{3}\mathcal{G}$ are defined as
\begin{splitequation}
	\mathcal{G}_{2}(\Phi_{v}) &= (\nabla_{ij}\Phi_{v})^{2} - (\nabla^{2}\Phi_{v})^{2} \\
	\Delta_{3}\mathcal{G} &= \mathcal{G}_{2}(\Phi) - \mathcal{G}_{2}(\Phi_{v}) \ ,
\end{splitequation}
where $\Phi$ and $\Phi_{v}$ refer to the normalised matter and velocity potentials, respectively.
Our bias model has thus the free parameters $b_{1}$, $b_{2}$, $\gamma_{2}$, and $\gamma_{3}^{-}$.
Following S17, we fix $\gamma_{2}$ to the local Lagrangian bias $\gamma_{2} = -\frac{2}{7}(b_{1}-1)$, which leaves us with three bias parameters per redshift bin.

The RSD power spectrum is modelled as \citep{Scoccimarro2004,Taruya2010}:
\begin{splitequation}
\label{equ:rsd-pofk}
	P(k,\mu) = W_{\infty}(i f k \mu)\left(P^{(1)}_{\rm novir}(k, \mu)+P^{(2)}_{\rm novir}(k, \mu)+P^{(3)}_{\rm novir}(k, \mu)\right)\,,\!\!\!
\end{splitequation}
where $f$ denotes the logarithmic growth rate and the generating function of the velocity differences in the large-scale limit $W_{\infty}(\lambda)$ includes non-linear corrections to account for the fingers-of-God effect and is parameterised in S17 as 
\begin{splitequation}
\label{equ:fog}
	W_{\infty}(\lambda) = \frac{1}{\sqrt{1-\lambda^{2}a_\mathrm{vir}^{2}}} \exp\left(\frac{\lambda^{2}\sigma_{v}^{2}}{1-\lambda^{2}a_\mathrm{vir}^{2}}\right) \ .
\end{splitequation}
Here $\sigma_{v}^{2}$ is given by $\sigma_{v}^{2} = \frac{1}{3}\int \mathrm{d}^{3}kP(k)/k^{2}$.
The velocity dispersion and higher moments of the velocity difference distribution, such as the kurtosis, are characterised by $a_\mathrm{vir}$, a free parameter that describes the contribution of velocities at small scales.
The $P_{\rm novir}$ terms in the bracket of Eq.~\eqref{equ:rsd-pofk} are computed using gRPT at one-loop order and the bias model of Eq.~\eqref{equ:bias-model} (see Sect.~3.1 and Appendix A in S17 for details).

The Alcock-Paczynski effect \citep{Alcock1979} is accounted for by rescaling $s=s' q(\mu')$ and $\mu=\mu'\frac{q_\parallel}{q(\mu')}$, where $q(\mu)=\sqrt{q_\parallel^2\mu'^{2} + q_\perp^2(1-\mu'^2)}$.
Here, $q_\perp = D_\mathrm{M}(z)/D_\mathrm{M}^\mathrm{fid}(z)$ and $q_\parallel = H^\mathrm{fid}(z) / H(z)$, where $D_\mathrm{M}(z)$ is the comoving angular diameter distance at the the mean redshift $z$ of the galaxy sample, $H(z)$ denotes the Hubble rate, and the superscript `fid' is assigned to quantities in the fiducial cosmology that was used to convert the measured redshifts to distances.

\subsubsection{Validation on simulation}
\label{sec:validation}
\begin{figure}
	\begin{center}
		\includegraphics[width=\columnwidth]{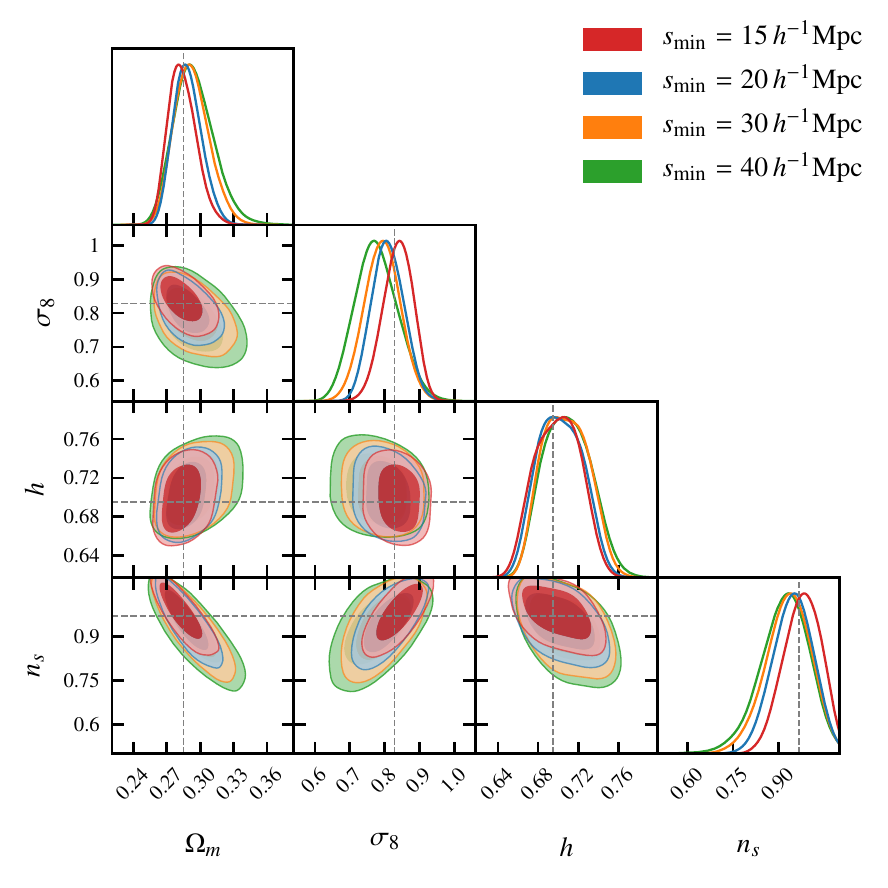}
		\caption{Cosmological parameters inferred from the \software{Minerva} mocks for different choices of the minimum separation scale used in the measurement. The true cosmology is indicated with dashed lines, while the cosmological constraints are shown in red ($s_{\rm min}=15\,h^{-1}\mathrm{Mpc}$), blue ($s_{\rm min}=20\,h^{-1}\mathrm{Mpc}$), orange ($s_{\rm min}=30\,h^{-1}\mathrm{Mpc}$), and green ($s_{\rm min}=40\,h^{-1}\mathrm{Mpc}$).}
		\label{fig:sim-scale-cuts}
	\end{center}
\end{figure}
The model described in Sect.~\ref{sec:model} has been extensively validated in S17. 
Further tests were done for the Fourier space wedges analysis of \citet{Grieb2017}, which used the same bias and RSD model.
The model was tested on the \software{Minerva} simulations \citep{Grieb2016,Lippich2019}, the BOSS RSD challenge, and the \software{MD-Patchy} mock catalogues \citep{Kitaura2016}. 

During these tests, the LSS parameters $q_\perp$, $q_\parallel$, and $f\sigma_8$ were varied. 
This parameter space does not map one-to-one to flat $\Lambda$CDM, since it allows to arbitrarily combine angular and radial distances, as well as the growth of structure. 
As we discuss in Sect.~\ref{sec:LSS-params}, restricting the sample space to flat $\Lambda$CDM can significantly tighten the parameter ranges allowed by the data. 

In light of this increased sensitivity, we deem it prudent to revisit some of the model validation carried out in S17. 
Specifically, we analyse the \software{Minerva} simulations using our RSD and bias model with the same parameters and priors as our cosmological results. 
The \software{Minerva} mocks were created from $N$-body simulations with $N=1000^3$ particles, evolved in a $L=1.5\,h^{-1}\mathrm{Gpc}$ box. 
The $z=0.31$ and $z=0.57$ snapshots were processed into a halo catalogue with a minimum halo mass of $M_\mathrm{min} = 2.67\times 10^{12}\,h^{-1}\mathrm{M}_\sun$ and then populated with the halo occupation distribution model of \citet{Zheng2007}. 

Figure~\ref{fig:sim-scale-cuts} shows the posteriors derived from the mean signal of 300 \software{Minerva} realisations, using a covariance matrix corresponding to one simulation volume. 
This yields parameter uncertainties that are at least 7\% smaller than those derived from the data. 
Figure~\ref{fig:sim-scale-cuts} demonstrates the effect of changing the minimum separation of the measurement on the inferred parameter constraints, analogous to figure 4 in S17. 
The input cosmology is recovered well for all scale cuts considered ($s_{\rm min}=15,\,20,\,30,\,40\,h^{-1}\mathrm{Mpc}$), consistent with the results of S17. 
While Fig.~\ref{fig:sim-scale-cuts} suggests that the model is robust down to a minimum separation of $s_{\rm min}=15\,h^{-1}\mathrm{Mpc}$, we nevertheless follow S17 with a minimum separation of $s_{\rm min}=20\,h^{-1}\mathrm{Mpc}$. 
Further tests on simulations are presented in Appendix \ref{sec:mocktests2}.

\subsubsection{Sampling and priors}
The parameter inference is performed with two pipelines: \software{CosmoMC} \citep{Lewis2002} -- the same setup as in S17 -- and \software{CosmoSIS} \citep{Zuntz2015}, using \software{MultiNest} \citep{Feroz2009, Feroz2013} to perform nested sampling.
The agreement with Planck is assessed using the public nuisance parameter-marginalised \software{plik\_lite\_TTTEEE}+\software{lowl}+\software{lowE} likelihood \citep{Planck2018-likelihood}.

For our fiducial analysis, we choose uninformative priors for all parameters except for $\Omega_\mathrm{b}h^2$, since BOSS is not able to constrain this parameter by itself. 
Even though our $\Omega_\mathrm{b}h^2$ prior is informative in the sense that it restricts the posterior, it is still chosen very conservatively, being approximately 25 times wider than the Planck uncertainty and $\sim10$ times wider than the recent big bang nucleosynthesis (BBN) constraints on $\Omega_\mathrm{b}h^2$ of \citet{Cooke2018}. 
Furthermore, we find that different $\Omega_\mathrm{b}h^2$ priors choices only impact the $h$ constraints, while leaving the other parameters virtually unchanged.
The upper prior ranges for $\Omega_\mathrm{c}h^2$ and $n_\mathrm{s}$ were lowered from those chosen in S17 to avoid numerical convergence issues, but remain uninformative.
Since the prior ranges for the non-linear bias and RSD parameters in S17 were restricting the posteriors, we significantly extend the prior ranges of these parameters in this analysis.

Our main cosmological parameter constraints are presented in Table~\ref{tab:fidparams}, while constraints from other prior choices and details of the sampled parameters and their priors are discussed in Appendix~\ref{sec:priors}.

\begin{table}
    \caption{Posterior constraints (marginal means with 68\% confidence interval) derived from BOSS DR12 data alone, as well as the combination of BOSS DR12 and cosmic shear from the Kilo-Degree Survey (KV450).}
    \begin{center}
        \begin{tabular}{lll}
        		\toprule
		Parameter             & BOSS       & BOSS+KV450   \\
		\midrule
$\Omega_\mathrm{m}                 $ & $0.317^{+0.015}_{-0.019}  $ & $0.323^{+0.014}_{-0.017}  $\\
$\sigma_8                 $ & $0.710\pm 0.049           $ & $0.702\pm 0.029           $\\
$h                        $ & $0.704\pm 0.024           $ & $0.691\pm 0.023           $\\
$n_s                      $ & $0.815\pm 0.085           $ & $0.863\pm 0.071           $\\
$S_8                      $ & $0.729\pm 0.048           $ & $0.728\pm 0.026           $\\                
	\bottomrule
        \end{tabular}
    \end{center}
    \label{tab:fidparams}
\end{table}%

\section{Results}
\label{sec:results}

\begin{figure*}
	\begin{center}
		\includegraphics[width=\textwidth]{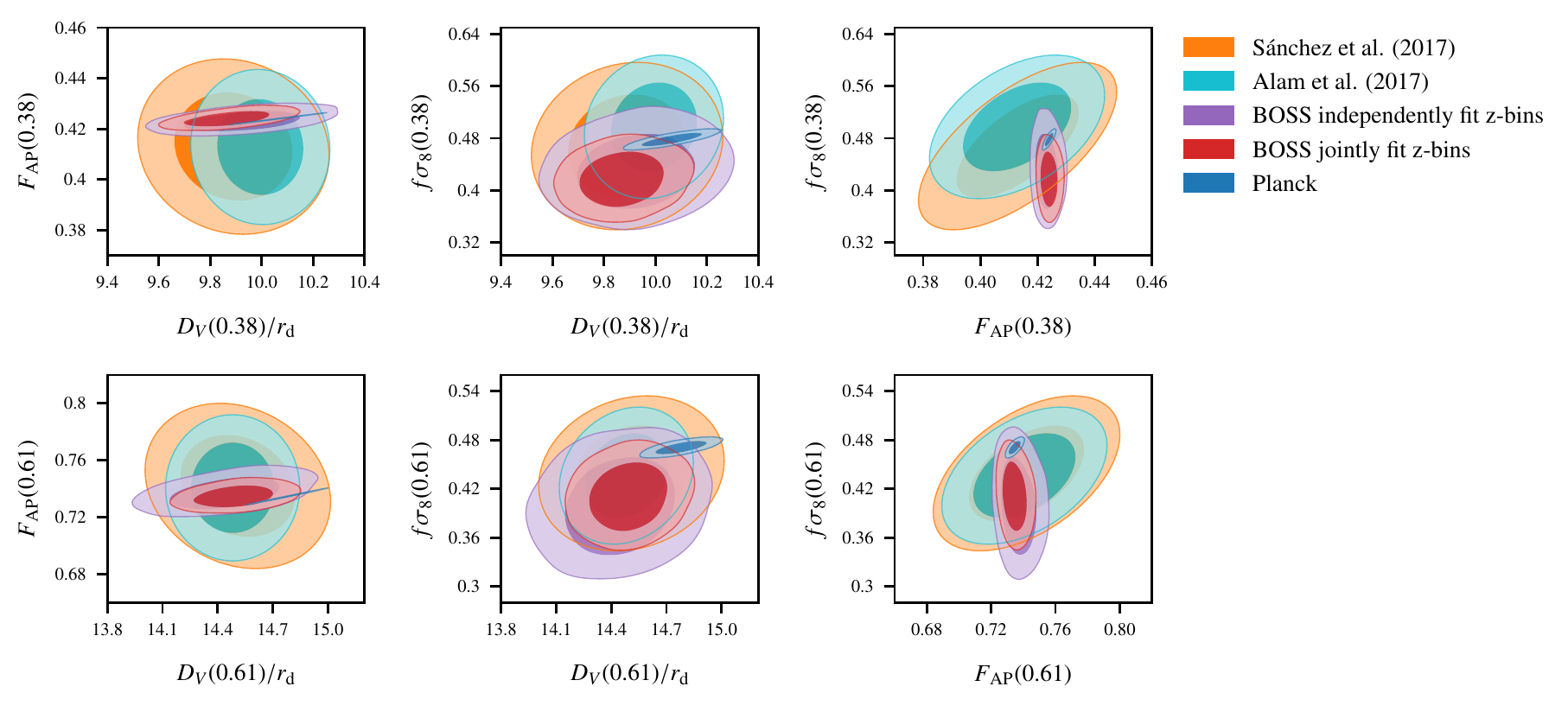}
		\caption{Constraints on the parameters $F_\mathrm{AP}$, $D_\mathrm{V}/r_{d}$, and $f\sigma_8$ at redshifts $z=0.38$ and $z=0.61$. The results from \citet{Sanchez2017} and the BOSS DR12 consensus analysis \citep{Alam2017} are shown in orange and cyan, respectively. Restricting the parameter space to flat $\Lambda$CDM in each BOSS redshift bin yields the purple contours. The joint constraints from both redshift bins (while sampling in flat $\Lambda$CDM) are shown in red. Finally, the blue contours correspond to the Planck 2018 constraints on these parameters.}
		\label{fig:LSS-params}
	\end{center}
\end{figure*}

\begin{figure}
	\begin{center}
		\includegraphics[width=\columnwidth]{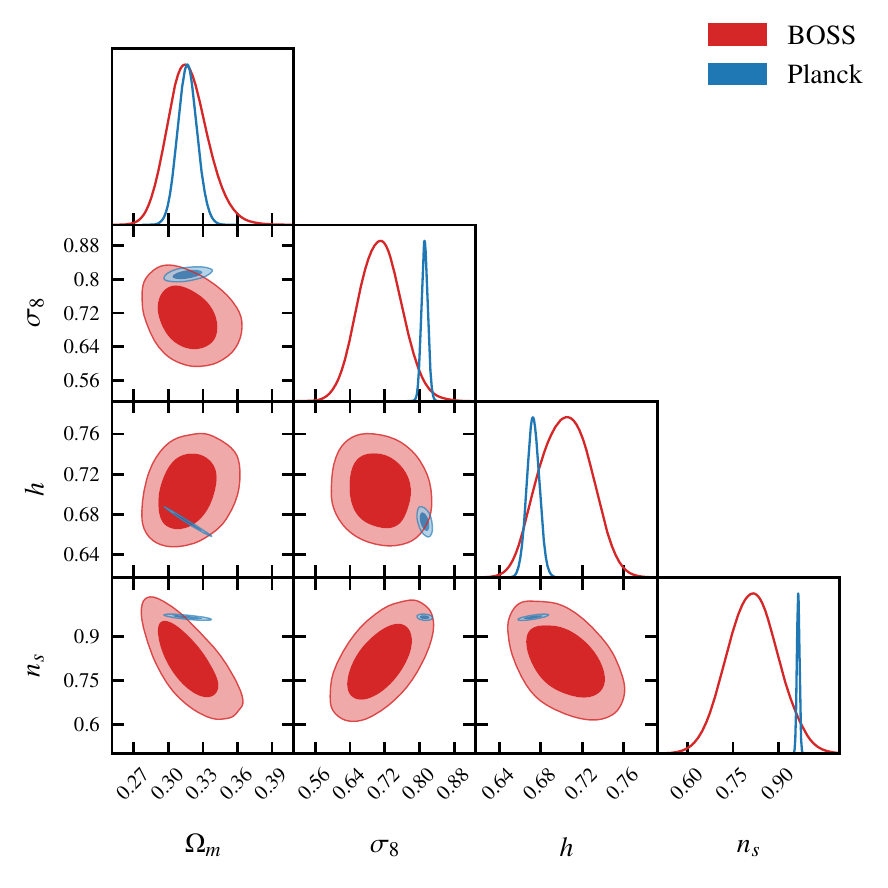}
		\caption{Constraints on flat $\Lambda$CDM derived from BOSS DR12 correlation function wedges (red) and Planck 2018 (blue).}
		\label{fig:cosmology-params}
	\end{center}
\end{figure}

\subsection{Constraining LSS}
\label{sec:LSS-params}
The BOSS DR12 consensus analysis \citep{Alam2017} does not constrain $\Lambda$CDM directly but rather the parameters $F_\mathrm{AP}(z)=D_\mathrm{M}(z)H(z)$, $D_\mathrm{V}(z) /r_{d}=\left(D_\mathrm{M}(z)^{2}cz/H(z)\right)^{\frac{1}{3}} / r_{d}$, and $f\sigma_8$, where $r_\mathrm{d}$ is the sound horizon at the drag epoch. 
In Fig.~\ref{fig:LSS-params} we present our constraints on these parameters at the mean redshifts $z=0.38$ and $z=0.61$ of the two redshift bins.
We consider two cases: first, we derive constraints individually for the two redshift bins, analogously to the BOSS analyses.
These individual constraints are shown in purple, while those from previous BOSS DR12 analyses are shown in orange (S17) and cyan \citep[BOSS DR12 consensus results,][]{Alam2017}, while the Planck 2018 results are in blue. 
Our constraints are in good agreement with those of S17 but are markedly tighter owing to the restrictions of the flat $\Lambda$CDM parameter space.
This shrinking of the allowed parameter range is especially pronounced for $F_\mathrm{AP}$ and can be understood by noting the tight correlation between $D_\mathrm{M}(z)$ and $H(z)$ in $\Lambda$CDM.
This correlation was not respected in S17, since there the shape of the linear power spectrum was fixed, while $q_\perp\frac{r_\mathrm{d}^\mathrm{fid}}{r_\mathrm{d}}$, $q_\parallel\frac{r_\mathrm{d}^\mathrm{fid}}{r_\mathrm{d}}$, and $f\sigma_8$ were varied.
The constraints can be further tightened by jointly analysing the two redshift bins, as is demonstrated by the red contours.

\subsection{Constraining $\Lambda$CDM}
Having established consistency with previous BOSS results and explored the increased sensitivity when restricting ourselves to flat $\Lambda$CDM, we now present the corresponding cosmological parameters.
Figure~\ref{fig:cosmology-params} presents the main results of this work; it shows the posterior distributions of $\Omega_\mathrm{m}$, the amount of matter in the Universe; $\sigma_8$, the present-day standard deviation of linear matter fluctuations on the scale of 8$\,h^{-1}\mathrm{Mpc}$; the Hubble parameter $h$; and the scalar power-law index $n_S$ for BOSS in red and Planck in blue. 
We find good agreement between BOSS and Planck, with $\sigma_{8}$ being the most deviant parameter, being low at $2.1\sigma$ significance. 
We also demonstrate internal consistency of our results: in Appendix~\ref{sec:lowz-vs-highz} we find consistency between the constraints from the two BOSS redshift bins analysed independently. 
In Appendix~\ref{sec:smax} we find consistent results when we reduce the maximum allowed clustering scale, removing large-scale data that is potentially biased by variations in the stellar density \citep{Ross2017}.
The posterior distributions for all sampled parameters are shown in Appendix~\ref{sec:priors}.

\subsection{Consistency with Planck}
In light of the low $\sigma_{8}$ values favoured by BOSS  we wish to quantify the agreement between BOSS and Planck over the whole parameter space. 
We consider two statistics: the Bayes' factor $R$, expressed as the ratio
\begin{equation}
    R=\frac{\mathcal{Z}_{\mathrm{BOSS+Planck}}}{\mathcal{Z}_\mathrm{BOSS} \mathcal{Z}_\mathrm{Planck}}
\end{equation}
between the evidence $\mathcal{Z}_{\text{BOSS+Planck}}$ for a model where the cosmological parameters are shared between BOSS and Planck, and the evidences $\mathcal{Z}_\mathrm{BOSS}$ and $\mathcal{Z}_\mathrm{Planck}$ for a model with separate sets of cosmological parameters.
\citet{Handley2019} pointed out the prior-dependence of the $R$ statistic and proposed a new statistic $S$, called `suspiciousness', that ameliorates the effect of the prior on the estimate of consistency.
Both statistics are computed using \software{anesthetic} \citep{anesthetic}. 

We find $\log R = 4.0\pm0.2$, corresponding to odds of {$57\pm13$} in favour of a single cosmology describing both BOSS and Planck. 
The suspiciousness is $\log S= 0.13\pm0.11$ with model dimensionality $d = 4.8\pm0.5$, which can be converted into a tension probability of $p=0.45\pm0.03$.
In terms of `sigmas', this corresponds to a {$0.76\pm0.05\sigma$} tension, indicating good agreement between BOSS and Planck.

\section{Discussion}
\label{sec:discussion}
In the previous section we have presented constraints on flat $\Lambda$CDM from the clustering of BOSS DR12 galaxies.
Our results agree with those of \citet{Loureiro2019}, who considered the angular power spectrum of BOSS DR12 galaxies in tomographic bins. 
Their parameter uncertainties are significantly larger than ours, however, owing to the restriction to large scales of their analysis.

Two recent analyses \citep{dAmico2019,Ivanov2019} of the BOSS DR12 power spectrum multipoles from \citet{Beutler2017} also found cosmological constraints very similar to ours.
Both analyses report a low amplitude of matter fluctuations compared to Planck: \citet{dAmico2019} find $\ln 10^{10} A_{s}=2.72\pm0.13$, while \citet{Ivanov2019} quote $\sigma_{8}= 0.721\pm 0.043$, both in excellent agreement with our results of $\ln 10^{10} A_{s}=2.74\pm 0.17$ and $\sigma_{8}=0.710\pm 0.049$.
Unlike our analysis, both \citet{dAmico2019} and \citet{Ivanov2019} fix $n_{s}$, and either fix the baryon fraction or impose a tight prior on $\Omega_\mathrm{b}h^2$.
Their theoretical modelling differs significantly to that of the present analysis, both in the treatment of matter clustering and, more importantly, that of RSD. 
Here we use a full parametric function for the fingers-of-God effect, Eq.~\eqref{equ:fog}, while they account for RSD (and other effects) by including a set of counter-terms. 
\citet{Ivanov2019} use the same biasing parametrisation as here, albeit with different priors. 
Nevertheless, the cosmological constraints are very similar between our analyses, signalling that the conclusions are not driven by improvements or changes in the theoretical model but by the BOSS data itself. 

While our results are consistent with Planck when considering the whole parameter space, the preference for low values of $\sigma_{8}$ is interesting in the context of other low-redshift cosmological probes, such as weak gravitational lensing. 
Weak lensing is sensitive to the parameter combination $S_{8} = \sigma_{8}\sqrt{\Omega_\mathrm{m}/0.3}$, which is found to be lower than that of Planck by all stage-3 weak lensing surveys \citep[e.g.,][]{Troxel2018, Hildebrandt2018, HSC-Cl}.

If there is new physics that affects the clustering of matter at low redshift relative to what one might expect based on CMB physics, it would be worthwhile to ask how we can combine low-redshift data sets to detect such new physics.
It has been shown that combining two-point statistics of gravitational lensing, galaxy positions, and their cross-correlations in so-called 3$\times$2pt analyses can yield powerful constraints on cosmology \citep{van-Uitert2018, Joudaki2018, DES-3x2pt}. 
These analyses did not make use of the full power of BOSS, however. 
While a full 3$\times$2pt analysis of BOSS and weak lensing would be beyond the scope of this \textit{Letter}, we showcase the potential of such a combination by considering a joint analysis of the results presented in Sect.~\ref{sec:results} with cosmic shear measurements from 450 sq. degrees of the optical and near-infra-red Kilo-Degree Survey \citep[KV450,][]{Hildebrandt2018}.   
We chose KV450 for convenience, but a similar analysis could also be carried out for weak lensing from the Dark Energy Survey \citep[DES,][]{Troxel2018} or Hyper Suprime-Cam \citep[HSC,][]{HSC-Cl,HSC-xi}.

\subsection{Joint analysis with weak lensing}
\begin{figure}
	\begin{center}
		\includegraphics[width=\columnwidth]{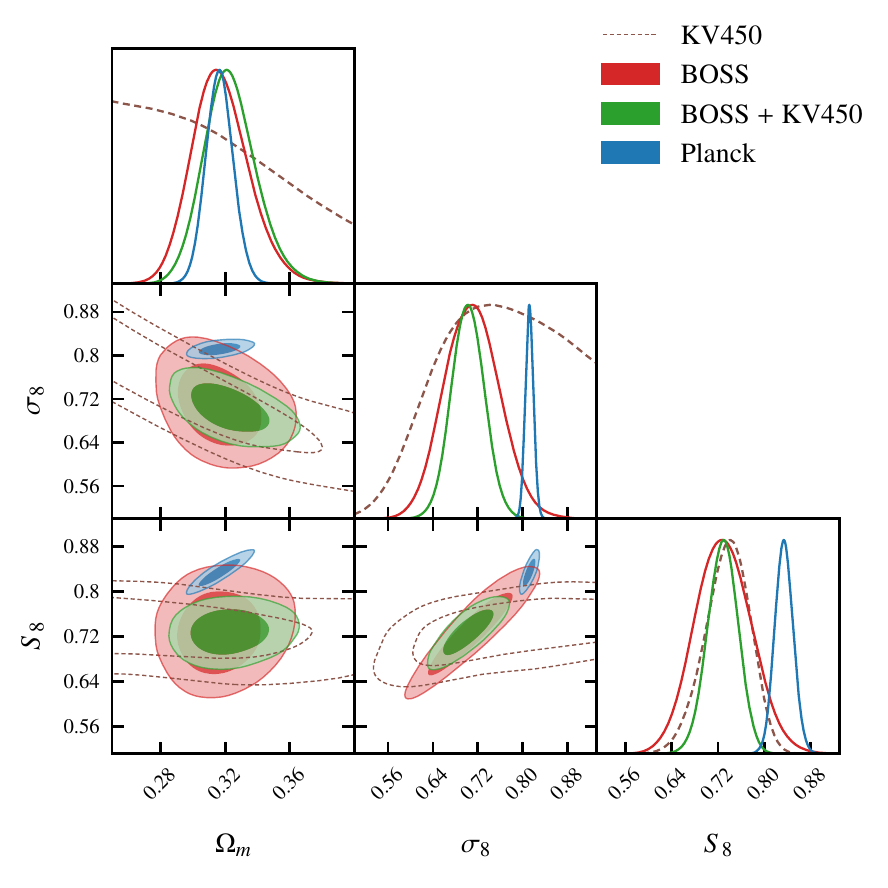}
		\caption{Constraints on flat $\Lambda$CDM when combining BOSS DR12 with KV450 (green). The red and blue contours are the same as Fig.~\ref{fig:cosmology-params} and denote the constraints from BOSS and Planck alone. The constraints from KV450 are shown with dashed lines. 
		}
		\label{fig:BOSS-KV450-cosmology-params}
	\end{center}
\end{figure}

Since the overlap region of the KV450 and BOSS footprints only account for 2\% of the BOSS area, we assume the two data sets to be independent. 
Inference can thus be carried out by simply multiplying the likelihoods. 
We take the \software{CosmoSIS} implementation of the KV450 likelihood, including all nuisance parameters, and add the bias and RSD model described in Sect.~\ref{sec:model}.
The resulting cosmology constraints are shown in Fig.~\ref{fig:BOSS-KV450-cosmology-params}. 
The BOSS-only and Planck contours are again shown in red and blue, respectively, while the joint constraints of BOSS and KV450 are in green. 
The KV450-only constraints are shown with dashed lines for illustrative purposes, as the priors, which are those used in \citet{Hildebrandt2018}, differ from those used for the other contours. 
There is excellent agreement on $S_{8}$ between BOSS and KV450 and the joint constraint of the two is $S_8 = 0.728\pm 0.026$, which is $3.4\sigma$ lower than Planck. 
The disagreement on $\sigma_{8}$ is even stronger, with BOSS and KV450 finding $\sigma_8 = 0.702\pm 0.029$, which is in $3.6\sigma$ tension with Planck. 
Over the whole parameter space, the odds in favour of a single cosmology describing the low and high-redshift Universe are {$7\pm2$} based on the Bayes factor, while the suspiciousness statistic $S$ indicates a {$2.1\pm0.3\sigma$} tension.

The value of $S_8$ measured by KV450 is consistent with, but lower than that of the DES and HSC collaborations. 
A joint analysis of BOSS with DES or HSC is therefore expected to be in less tension with Planck than the joint BOSS and KV450 analysis presented here. 
We note however that different methodologies have been used to estimate the redshift distribution of source galaxies. 
Adopting a consistent treatment results in an even better agreement between KV450 and DES \citep{Joudaki2019, Asgari2019}. 

\section{Conclusions}
\label{sec:conclusions}

In this \textit{Letter} we have shown that the clustering of BOSS DR12 galaxies can constrain flat $\Lambda$CDM without relying on other data sets.
Anisotropic galaxy clustering measurements thus provide a new tool to independently probe the cosmology of the low-redshift Universe. 
Data from future redshift surveys such as the Dark Energy Spectroscopic Instrument \citep[][]{DESI-Collaboration2016}, will further increase the power of the analysis presented in this work, and in conjunction with other low-redshift probes, provide a powerful complement to cosmology derived from CMB observations. 

We restricted ourselves to flat $\Lambda$CDM in the present analysis.
Relaxing this assumption and considering cosmologies that allow for curvature, varying masses of the neutrinos, or extensions beyond $\Lambda$CDM severely degrades the constraining power of Planck and makes it reliant on other data, such as galaxy clustering, to break parameter degeneracies \citep{Planck2018-cosmology}. 
In light of the findings of this \textit{Letter}, it is then intriguing to ask if and how well BOSS can constrain these extended cosmologies by itself. 
We will consider such analyses in forthcoming work.

\begin{acknowledgements}
We thank Shadab Alam, Roman Scoccimarro, and Joe Zuntz for useful discussions.
The figures in this work were created with \software{matplotlib} \citep{Hunter2007} and \software{getdist}, making use of the 
\software{numpy} \citep{Oliphant2006} and \software{scipy} \citep{Jones2001} software packages. 

TT acknowledges funding from the European Union’s Horizon 2020 research and innovation programme under the Marie Sk{l}odowska-Curie grant agreement No 797794.
AGS acknowledges support by the German Research Foundation cluster of excellence ORIGINS (EXC 2094, www.origins-cluster.de).
We acknowledge support from the European Research Council under grant numbers 647112 (MA, CH, CL), 770935 (HH, AW), and 693024 (SJ).
CH also acknowledges support from the Max Planck Society and the Alexander von Humboldt Foundation in the framework of the Max Planck-Humboldt Research Award endowed by the Federal Ministry of Education and Research.

HH also acknowledges support from a Heisenberg grant of the Deutsche Forschungsgemeinschaft (Hi 1495/5-1).

SJ also acknowledges support from the Beecroft Trust.

AK acknowledges support from Vici grant 639.043.512, financed by the Netherlands Organisation for Scientific Research (NWO).

Funding for SDSS-III has been provided by the Alfred P. Sloan Foundation, the Participating Institutions, the National Science Foundation, and the U.S. Department of Energy Office of Science. The SDSS-III web site is http://www.sdss3.org/.

SDSS-III is managed by the Astrophysical Research Consortium for the Participating Institutions of the SDSS-III Collaboration including the University of Arizona, the Brazilian Participation Group, Brookhaven National Laboratory, Carnegie Mellon University, University of Florida, the French Participation Group, the German Participation Group, Harvard University, the Instituto de Astrofisica de Canarias, the Michigan State/Notre Dame/JINA Participation Group, Johns Hopkins University, Lawrence Berkeley National Laboratory, Max Planck Institute for Astrophysics, Max Planck Institute for Extraterrestrial Physics, New Mexico State University, New York University, Ohio State University, Pennsylvania State University, University of Portsmouth, Princeton University, the Spanish Participation Group, University of Tokyo, University of Utah, Vanderbilt University, University of Virginia, University of Washington, and Yale University.

Based on data products from observations made with ESO Telescopes at the La Silla Paranal Observatory under programme IDs 177.A-3016, 177.A-3017 and 177.A-3018. 

\end{acknowledgements}

\bibliographystyle{aa}
\bibliography{references.bib}

\begin{appendix}

\section{Validation on simulations}
\label{sec:mocktests2}
\begin{figure}
	\begin{center}
		\includegraphics[width=\columnwidth]{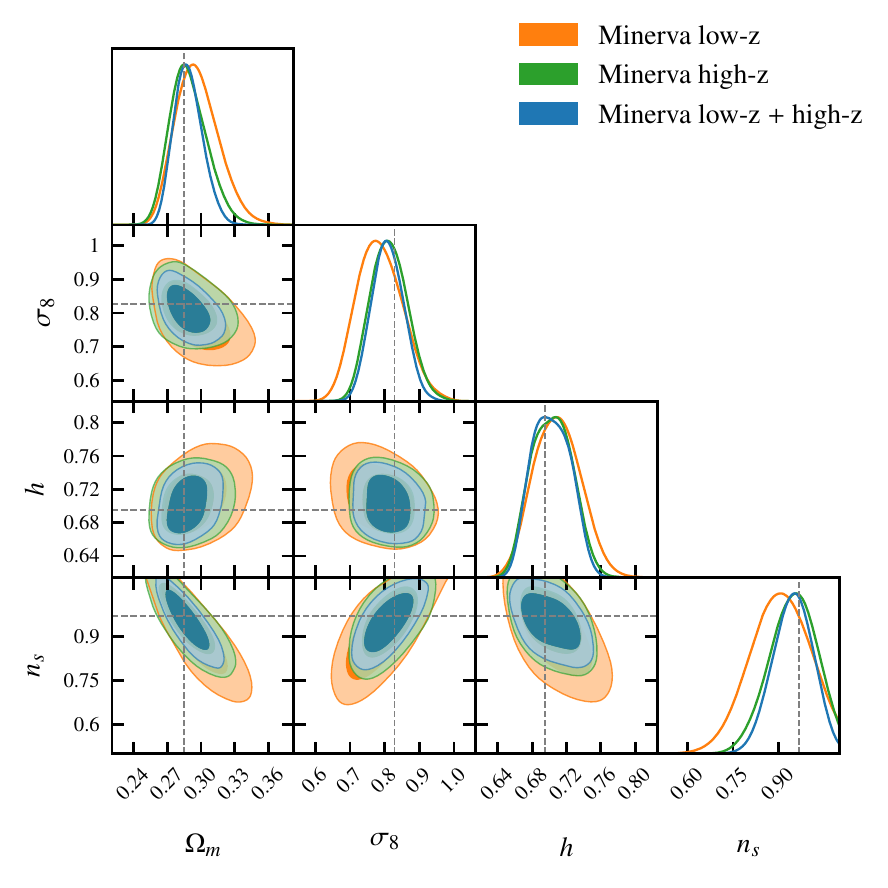}
		\caption{Cosmological parameters inferred from the \software{Minerva} mocks for low-redshift sample (orange), high-redshift sample (green), and joint analysis of both samples (blue). The true cosmology is indicated with dashed lines.}
		\label{fig:minerva-lowz-vs-highz}
	\end{center}
\end{figure}

The validation tests described in Sect.~\ref{sec:validation} consider a mock galaxy population that resembles the low and high-redshift bins of the combined DR12 galaxy sample. 
Beside checking the effect of the minimum physical scale on the parameter constraints for the joint-analysis of the low and high-redshift bins (see Fig.~\ref{fig:sim-scale-cuts}), we also perform this test for the two bins individually. 
We find both bins to be robust against changes of the minimum separation $s_\mathrm{min}$. 
The parameter constraints from the mocks for $s_\mathrm{min}=20\,h^{-1}\mathrm{Mpc}$ for the two redshift bins individually, as well as combined, is shown in Fig.~\ref{fig:minerva-lowz-vs-highz}.

\section{Parameter constraints and prior choices}
\label{sec:priors}
\begin{figure*}
	\begin{center}
		\includegraphics[width=\textwidth]{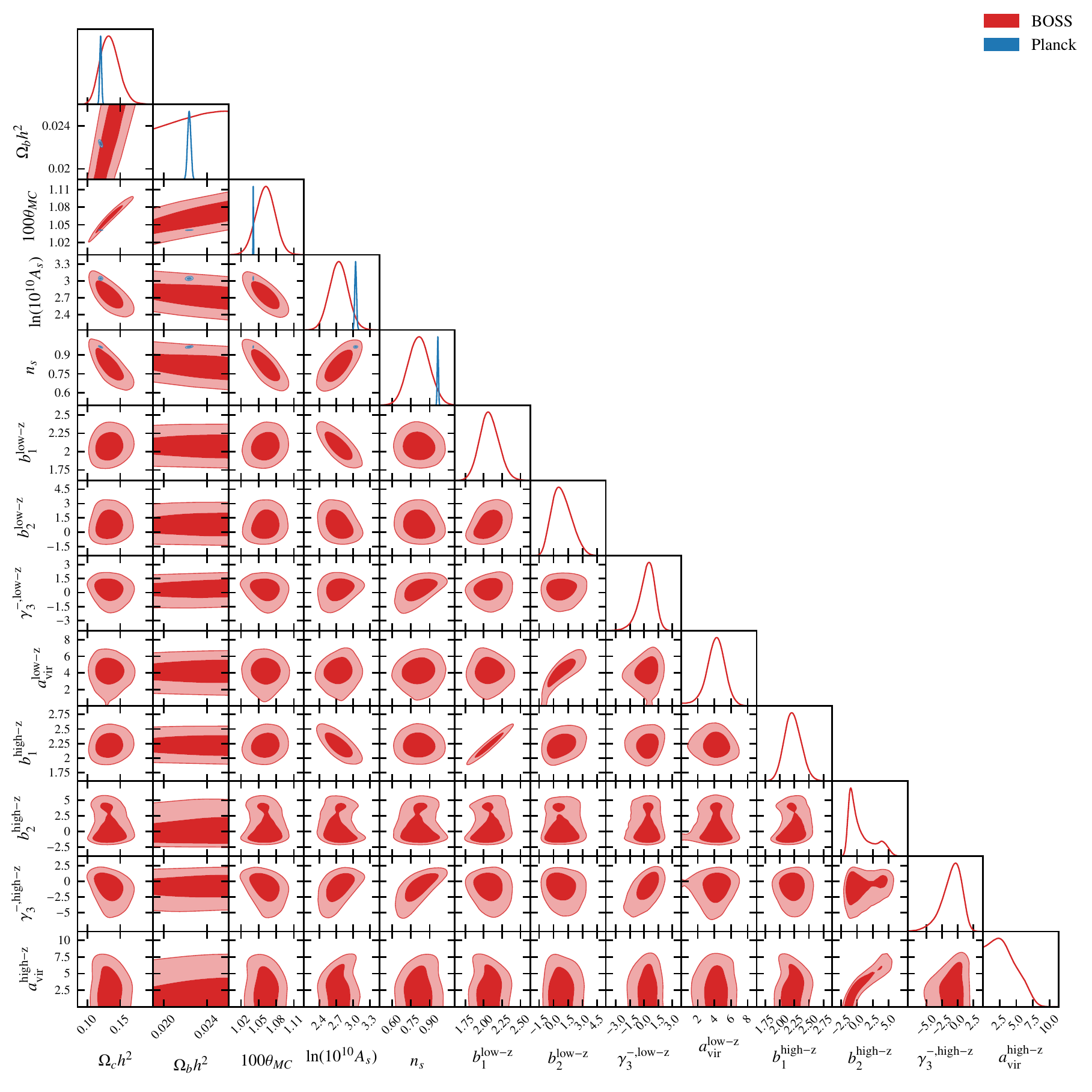}
		\caption{Posterior distributions for all sampled parameters in our main analysis. The posteriors derived from the BOSS clustering wedges are shown in red, while those from Planck 2018 are shown in blue.}
		\label{fig:all-params}
	\end{center}
\end{figure*}

The sampled parameters, their priors, and marginal posteriors for the fiducial analysis are listed in Table~\ref{tab:params-full}. 
We furthermore assume a single massive neutrino with a mass of $0.06\,\mathrm{eV}$. 
Figure~\ref{fig:all-params} shows the posterior distributions of all sampled parameters of our model, consisting of five cosmological parameters $\Omega_\mathrm{c}h^2$, $\Omega_\mathrm{b}h^2$, $100\theta_\mathrm{MC}$, $\ln 10^{10} A_{s}$, and $n_{s}$; and the bias and RSD parameters $b_{1}$, $b_{2}$, $\gamma^{-}_{3}$, and $a_\mathrm{vir}$ for each redshift bin. 
All parameters, except for $\Omega_\mathrm{b}h^2$, are constrained by the data. 
The RSD parameter $a_\mathrm{vir}$ can only take on positive values; the lack of a lower limit on $a_\mathrm{vir}$ of the high-z bin is therefore not an artefact of the prior choice.

The best-fit model has a $\chi^{2}$ of 172.1 for 168 data points, in agreement with S17. 

\begin{figure}
	\begin{center}
		\includegraphics[width=\columnwidth]{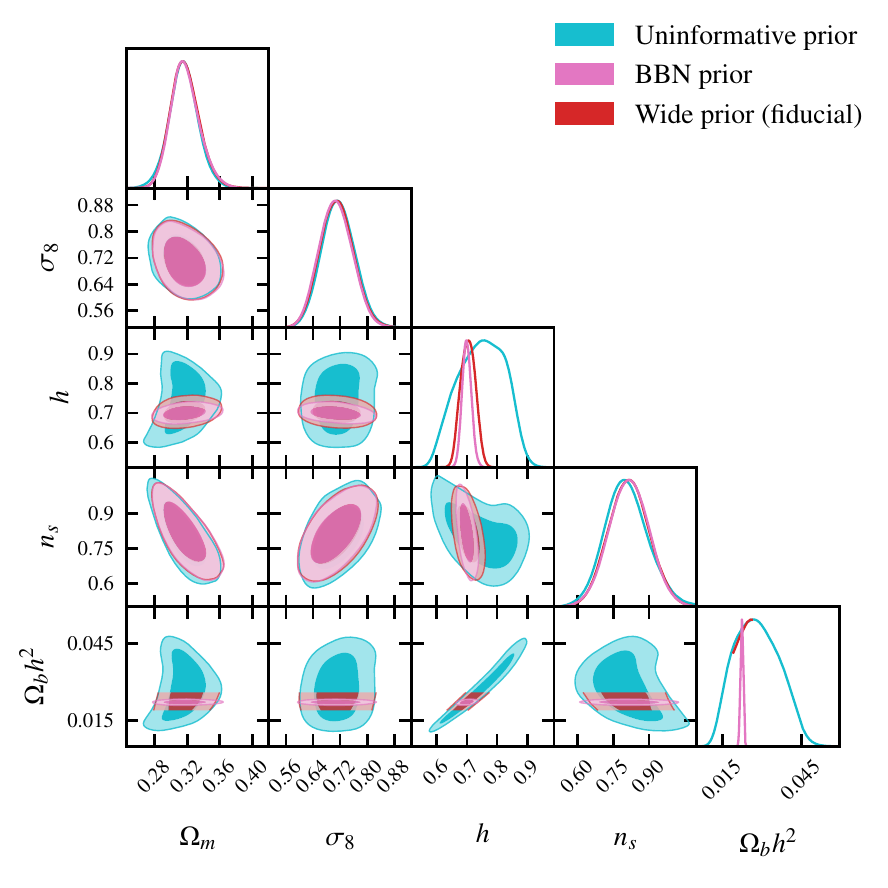}
		\caption{Constraints on flat $\Lambda$CDM derived from BOSS DR12 correlation function wedges for different prior choices for $\Omega_\mathrm{b}h^2$: an uninformative, flat prior (cyan), a BBN-based prior (pink), and the wide and flat prior used in the fiducial analysis (red).}
		\label{fig:cosmology-bbn-priors}
	\end{center}
\end{figure}

To quantify the impact of the choice for the $\Omega_\mathrm{b}h^2$ prior in our fiducial analysis, we also derive parameter constraints for the case where we allow $\Omega_\mathrm{b}h^2$ to very freely using a uniform prior between 0.005 and 0.1.
Using such an uninformative prior degrades the constraints on $h$ and $\theta_{\rm MC}$ due to their degeneracy with $\Omega_\mathrm{b}h^2$ but leaves the constraints on the other parameters unchanged. 
We also consider the case where we impose a BBN prior on $\Omega_\mathrm{b}h^2$. 
Specifically, we use the conservative BBN prior $\Omega_\mathrm{b}h^2 = 0.0222  \pm 0.0005$, which was derived in \citet{Planck2018-cosmology} based on the primordial deuterium abundance measurements of \citet{Cooke2018}. 
As with the uninformative $\Omega_\mathrm{b}h^2$ prior, only the constraints on $h$ are impacted, for which we find $h = 0.700\pm 0.015
$. 
The constraints for the three prior choices are shown in Fig.~\ref{fig:cosmology-bbn-priors}.

All fiducial chains were run until the \citet{Brooks1998} convergence criterion of $R-1<0.01$ was reached. 
Some of the ancillary chains had a slightly weaker convergence criteria but all chains achieved at least $R-1<0.02$.
	
\begin{table}
    \caption{Priors used in this work and in S17, as well as our posteriors (marginal means with 68\% confidence interval) derived from BOSS DR12 data alone. The priors on the cosmological parameters, as well as the bias and RSD parameters are all uniform (indicated by $U(\dots)$).}
    \begin{center}
        \begin{tabular}{llll}
        		\toprule
		Parameter             & Prior (S17)       & Prior (this work)    & BOSS  \\
		\midrule
		$\Omega_\mathrm{c} h^{2}$    & $U(0.01, 0.99)$   & $U(0.01, 0.2)$       & $0.134^{+0.012}_{-0.016}  $ \\
		$\Omega_\mathrm{b} h^{2}$    & $U(0.005, 0.1)$   & $U(0.019, 0.026)$    & ---                         \\
		$100\theta_{\rm MC}$  & $U(0.5, 10.0)$    & $U(0.5, 10.0)$       & $1.062\pm 0.016           $ \\
		$\ln 10^{10} A_\mathrm{s}$   & $U(2.0, 4.0)$     & $U(1.5, 4.0)$        & $2.74\pm 0.17             $ \\
		$n_{s}$               & $U(0.8, 1.2)$     & $U(0.5, 1.1)$        & $0.815\pm 0.085           $ \\
		\midrule                                                                                       
		Low-z                 &                   &                      &                             \\
		$b_{1}$               & $U(0.5, 9.0)$     & $U(0.5, 9.0)$        & $2.08^{+0.12}_{-0.14}     $ \\
		$b_{2}$               & $U(-4.0, 4.0)$    & $U(-4.0, 8.0)$       & $0.86^{+0.84}_{-1.2}      $ \\
		$\gamma^{-}_{3}$      & $U(-3.0, 3.0)$    & $U(-8.0, 8.0)$       & $0.29^{+0.95}_{-0.63}     $ \\
		$a_\mathrm{vir}$                 & $U(0.2, 5.0)$     & $U(0.0, 12.0)$       & $4.12^{+1.2}_{-0.96}      $ \\
		\midrule                                                                                       
		High-z                &                   &                      &                             \\
		$b_{1}$               & $U(0.5, 9.0)$     & $U(0.5, 9.0)$        & $2.22^{+0.13}_{-0.15}     $ \\
		$b_{2}$               & $U(-4.0, 4.0)$    & $U(-4.0, 8.0)$       & $0.66^{+0.71}_{-2.4}      $ \\
		$\gamma^{-}_{3}$      & $U(-3.0, 3.0)$    & $U(-8.0, 8.0)$       & $-1.0^{+1.9}_{-1.1}       $ \\
		$a_\mathrm{vir}$                 & $U(0.2, 5.0)$     & $U(0.0, 12.0)$       & $< 3.95                   $ \\
		\midrule                                                                                       
		$h$                   & ---               & ---                  & $0.704\pm 0.024           $ \\
		$\Omega_\mathrm{m}$          & ---               & ---                  & $0.317^{+0.015}_{-0.019}  $ \\
		$\sigma_{8}$          & ---               & ---                  & $0.710\pm 0.049           $ \\
		$S_{8}$               & ---               & ---                  & $0.729\pm 0.048           $ \\
		\bottomrule
        \end{tabular}
    \end{center}
    \label{tab:params-full}
\end{table}%

\section{Low-z and high-z}
\label{sec:lowz-vs-highz}
\begin{figure}
	\begin{center}
		\includegraphics[width=\columnwidth]{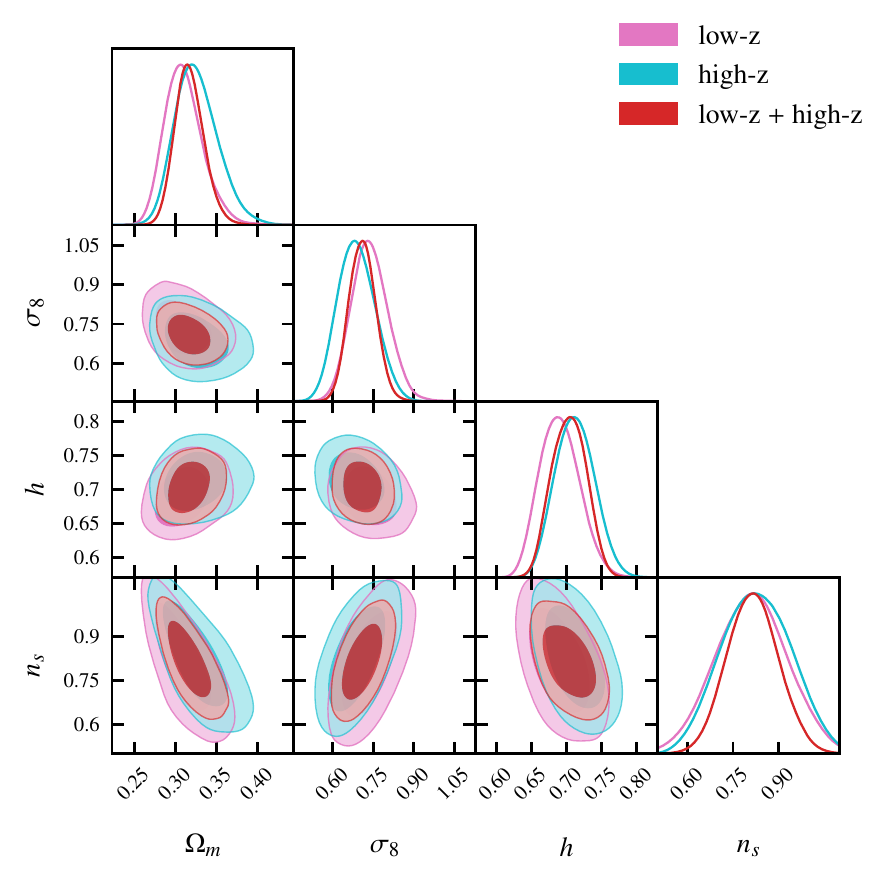}
		\caption{Constraints on flat $\Lambda$CDM derived from BOSS DR12 correlation function wedges using only the low-redshift bin (pink), only the high-redshift bin (cyan), and the joint constraints used in the main analysis (red).}
		\label{fig:cosmology-params-lowz-vs-highz}
	\end{center}
\end{figure}

Figure~\ref{fig:cosmology-params-lowz-vs-highz} presents the posterior distributions of $\Omega_\mathrm{m}$, $\sigma_{8}$, $h$, and $n_{s}$ analogously to Fig.~\ref{fig:cosmology-params} but considering the two redshift bins separately. 
We find that the two redshift bins yield consistent parameter constraints, considering that the two bins are independent. 
The small differences between the low- and high-redshift bins furthermore agree well with those found in \citet{Ivanov2019}.

\section{Dependence on $s_\mathrm{max}$}
\label{sec:smax}
\begin{figure}
	\begin{center}
		\includegraphics[width=\columnwidth]{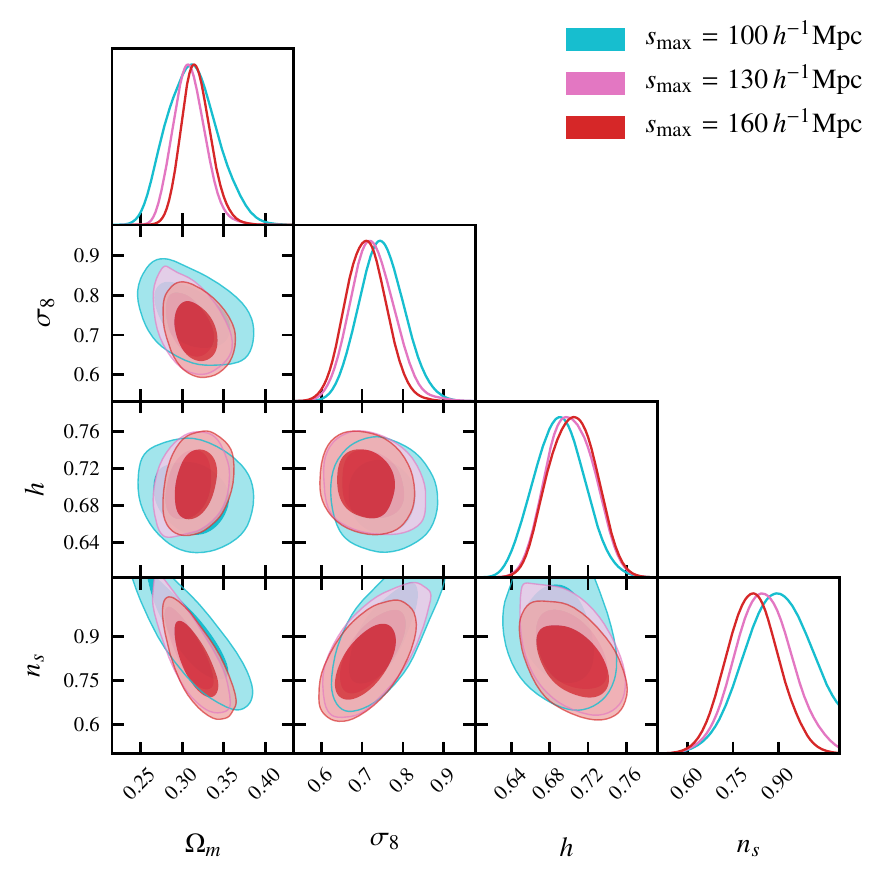}
		\caption{Posterior distribution of the cosmological parameters when restricting the maximum separation to $s_\mathrm{max}=100\,h^{-1}\mathrm{Mpc}$ (pink), $s_\mathrm{max}=130\,h^{-1}\mathrm{Mpc}$ (cyan), and the fiducial $s_\mathrm{max}=160\,h^{-1}\mathrm{Mpc}$ (red).}
		\label{fig:cosmology-params-smax}
	\end{center}
\end{figure}

Variations of the stellar density across the sky affect the selection function of BOSS DR12 galaxies and thus their clustering signal. 
\citet{Ross2017} showed that the weights assigned to the BOSS DR12 galaxies sufficiently mitigate this systematic for BAO measurements. 
In a full-shape analysis, such a residual systematic would boost the clustering signal at large scales, thus causing the data to prefer lower values of $n_{s}$.
To test for this possibility, we repeat the parameter inference but restrict the maximum separation to $s_\mathrm{max}=100\,h^{-1}\mathrm{Mpc}$ and $s_\mathrm{max}=130\,h^{-1}\mathrm{Mpc}$. 
The resulting posterior distributions are shown in Fig.~\ref{fig:cosmology-params-smax}. 
Both cuts yield consistent results with our fiducial choice of $s_\mathrm{max}=160\,h^{-1}\mathrm{Mpc}$, which was also employed in S17. 

\end{appendix}

\end{document}